# Heterogeneous confinement in laterally coupled InGaAs/GaAs quantum dot molecules under lateral electric fields

J. Peng,[1] C. Hermannstädter,[2] M. Witzany,[2] M. Heldmaier,[2] L. Wang,[3,1] S. Kiravittaya,[3] A. Rastelli,[3] O. G. Schmidt,[3] P. Michler,[2] and G. Bester[1]

[1]*Max-Planck-Institut für Festkörperforschung, Heisenbergstr. 1, D-70569 Stuttgart, Germany*
[2]*IHFG, Universität Stuttgart, Allmandring 3, D-70569 Stuttgart, Germany*
[3]*Institute for Integrative Nanosciences, IFW Dresden, Helmholtzstr. 20, D-01069 Dresden, Germany*

We study the electronic and optical properties of laterally coupled InGaAs/GaAs quantum dot molecules under lateral electric fields. We find that the electrons perceive the double-dot structure as a compound single object and tunnel through a basin connecting the dots from underneath. The holes discern two well-separated dots and are unable to tunnel. Through a combination of predictive atomistic modeling, detailed morphology studies, and single-object microphotoluminescence measurements, we show that this peculiar confinement results in an unusual heterogeneous behavior of electrons and holes with profound consequences on optical properties. We find a qualitatively different signal in optical-absorption, emission under resonant, and emission under nonresonant excitations. We explain this behavior by invoking the carriers' dynamics following light absorption.

## I. INTRODUCTION

Recently, the initially almost disconnected fields of solid-state physics and quantum optics have witnessed some stimulating interactions. These have been triggered by the availability of newly designed structures, such as quantum dots (QDs), thought to be able to provide the well-defined and isolated quantum levels at the heart of quantum optics. Some initial success of these interactions has to be acknowledged, such as the generation of single photons and entangled photon pairs from quantum dots[1–5] and first steps toward the coherent manipulation and preparation of quantum states.[6–8] However, it also becomes increasingly clear that the initially rather naive view on few level systems realized in condensed matter has to be realistically reassessed by higher-level quantitative theories of realistic, "as-grown," structures. The necessary initial step for such a work is the experimental determination of shape, size, and composition profiles that have to be fed into the theory. Laterally coupled quantum dot molecules (QDMs) are good candidates for such a study since their morphology has been investigated recently in detail.[9] They are relevant to the field of nanostructure-based quantum information science, where these structures are envisioned as building blocks.[10] While the theoretical work needs structural input from experiment, it also relies significantly on spectroscopic information. As will become clear in the following, it is the combination of detailed morphology studies, modeling, and spectroscopy that leads to a profound understanding of their electronic and optical properties.

We find an unusual situation, where the hole states are mainly uncoupled, leading to an eigenvalue spectrum akin the ones of two isolated quantum dots. The electron states, however, are electronically coupled and split into bonding and antibonding states. The coupling is favored by an In-rich region connecting both dots from underneath and the magnitude of the bonding-antibonding splitting for electrons, i.e., the electronic hopping term, depends on both the dots and this "basin," connecting the dots. The coupling is 1.1 meV when the two dots are touching each other and decreases to 0.6 meV when the two dots are 8 nm apart. The rather small magnitude of this splitting may be deceiving as it does not point to a weak coupling created by a large interdot barrier. On the contrary, the basin lays a barrier-free path connecting both dots. We may describe the electrons as perceiving the structure as a single object rather than two different dots. This electronic coupling is surprising from the point of view of the large dot-center separation of more than 40 nm. We show that this heterogeneous character of the structure (holes see two dots, electrons see one compound structure) leads to strong effects on the photoluminescence (PL), where the dynamical processes of the exciton formation comes prominently into play. Consequently, we obtain qualitatively different signals in absorption, in PL under resonant, and in PL under nonresonant excitation. Furthermore, the electron mediated tunneling and coupling are of particular interest in lateral QDMs in the context of quantum information science based on single electron and spin properties.

## II. GROWTH AND CHARACTERIZATION

The unique growth process, that combines molecular-beam epitaxy and *in situ* atomic-layer precise etching, leads to the creation of a low-density ensemble of InAs/GaAs double dots aligned along the $[1\bar{1}0]$ crystal axis.[10,11] During the growth of a QDM, first, a nanohole, referred to as basin in the following, is formed which is subsequently overgrown supporting the formation of two lens-shaped dots characterized in detail in Ref. 9. For optical spectroscopy, the QDMs were thermally treated during a growth interruption in order to blueshift their emission to 1.32–1.36 eV and were embedded in a planar cavity.[12] Figure 1 shows (a) a top view and (b) a cross-sectional view of the QDM structure used in the simulations. A lateral electric field is applied along the $[1\bar{1}0]$ direction (positive fields point from the left to the right dot).

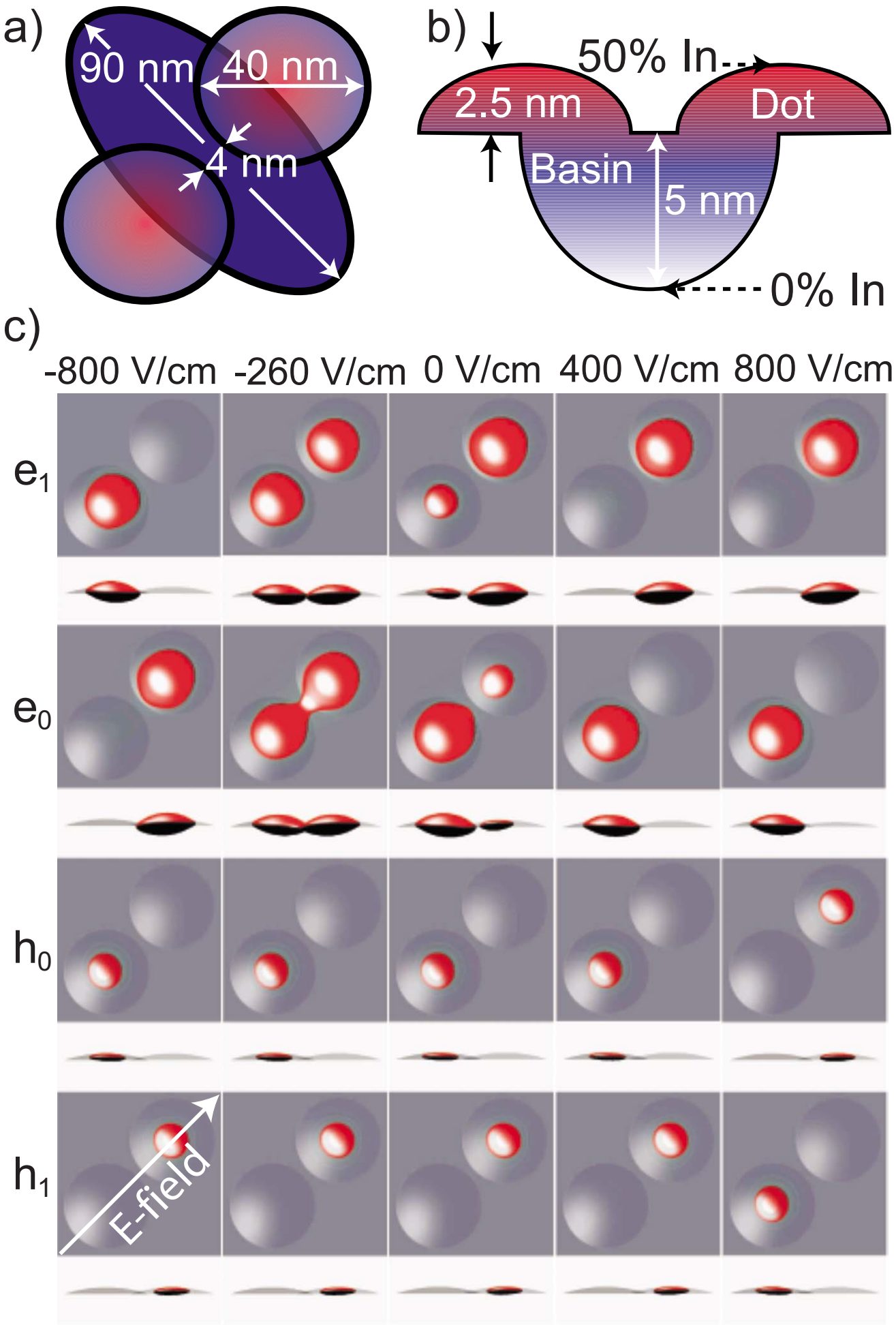

FIG. 1. (Color online) (a) Top view [(001) plane] and (b) cross-sectional view [(110) plane] of the two lens-shaped dots and the basin. (c) Square of the single-particle electron (hole) wave functions $e_{0,1}$ ($h_{0,1}$) as a function of the applied in-plane electric field. The shape of the dots is given in light gray and the isosurfaces contains 75% of the state densities. The applied field is within the range of −800 to 800 V/cm and the positive direction is shown in the figure.

The dimensions given in Fig. 1 have been used for most of the results presented in the figures. However, we varied the dot separation between 0 and 8 nm, the dot diameters between 38 and 44 nm, allowing for dots of different sizes, and simulated a QDM with dots of dissimilar heights. This variation in sizes follows the experimental characterization.[9]

## III. THEORETICAL METHOD

We use the atomistic empirical pseudopotential approach[13,14] to obtain the single-particle eigenvectors and eigenfunctions. This method takes strain, band coupling, coupling between different parts of the Brillouin zone, and spin-orbit coupling into account, retaining the atomistically resolved structure. Hence, two structures with identical overall shapes and average compositions $x$ are distinct since each structure is the product of a process where the uncommon atoms (In and Ga) occupy their sites randomly but keeping an overall concentration of $x$. The excitonic properties are calculated using the configuration-interaction approach.[15] A review of the method can be found in Ref. 14.

## IV. SPECTROSCOPY

Due to the low spatial density ($<10^8$ cm$^{-2}$), low-temperature (4.2 K) micro-PL measurements can be performed on single QDMs. The PL was dispersed using a 0.75 m spectrometer and detected with a liquid-nitrogen-cooled charge-coupled device. To date, we have studied more than 100 QDMs under the influence of lateral electric fields using resonant and nonresonant optical excitation with a Ti:sapphire laser. At low power excitation, two sharp lines separated by 0.3–2.5 meV, depending on the molecule, dominate the spectra. These two lines are attributed to direct neutral excitonic recombinations in either of the two dots. The relative intensities of these two spectral lines can be switched using a lateral electric field.[10,12] The detail of this coupling and tuning mechanism are discussed in the subsequent sections.

## V. ELECTRONIC COUPLING UNDER LATERAL ELECTRIC FIELD

In Fig. 1(c), we show the calculated single-particle electron ($e_0$,$e_1$) and hole ($h_0$,$h_1$) state probability densities for five different lateral electric fields applied along the $[1\bar{1}0]$ direction. For each state, a top view and a side view are shown. The electron states undergo a smooth transition where the electric field pulls the wave function from one dot to the other. The state $e_0$ represents a bonding state with some occupation probability in the coupling region while the state $e_1$ is antibonding. The hole states are more localized and abruptly change from one dot to the other. This behavior can be quantified by looking at the eigenvalues of the electron and hole states as a function of the electric field in Figs. 2(a) and 2(b). The dominant wave-function localization is described by "L" and "R" for left and right dots. For instance, for a high negative field, the first electron ($e_0$) and first hole ($h_0$) states are localized on the right and left dots, respectively. The electron states undergo an anticrossing with a 0.77 meV splitting at the tuning field $F_e$=−260 V/cm while the hole states cross at the field $F_h$=480 V/cm.

The experimental characterization of the structure given in Figs. 4(c) and 4(f) of Ref. 9 shows variations in sizes. The height of the QDM after capping [Ref. 9, Fig. 4(c)] shows a rather sharp distribution around 2.6 nm. However, the diameter [given in the inset of Ref. 9, Fig. 4(c)] shows a distribution between 40 and 48 nm. The dot separation [Ref. 9, Fig. 4(f)] varies between 0 and 12 nm. To survey the experimental ranges of possible morphologies given in Fig. 4 of Ref. 9 we have done the following calculations.

### A. Dependence on dot separation

We have calculated the full field dependence for the dot separation 0, 2, 4, 6, and 8 nm, where both dots have 40 nm diameter. We have plotted the bonding-antibonding coupling $\Delta_{BA}$ for electronic states [see Fig. 2(a)], i.e., the electronic hopping term, in Fig. 3(a). The electronic hopping term $\Delta_{BA}$ is reduced with the dot separation but goes to an asymptotic value different from 0. This is the consequence of the coupling being not a pure dot-dot barrier-tunneling process but partly occurring through the underlying basin. The solid red curve in Fig. 3(a) gives the best fit to the data points achieved using the exponential dependence of bonding-antibonding coupling (millielectron volt) on the dot separation $\delta$ (nanometer),

$$\Delta_{BA} = 0.503 e^{-0.275\delta} + 0.6.$$

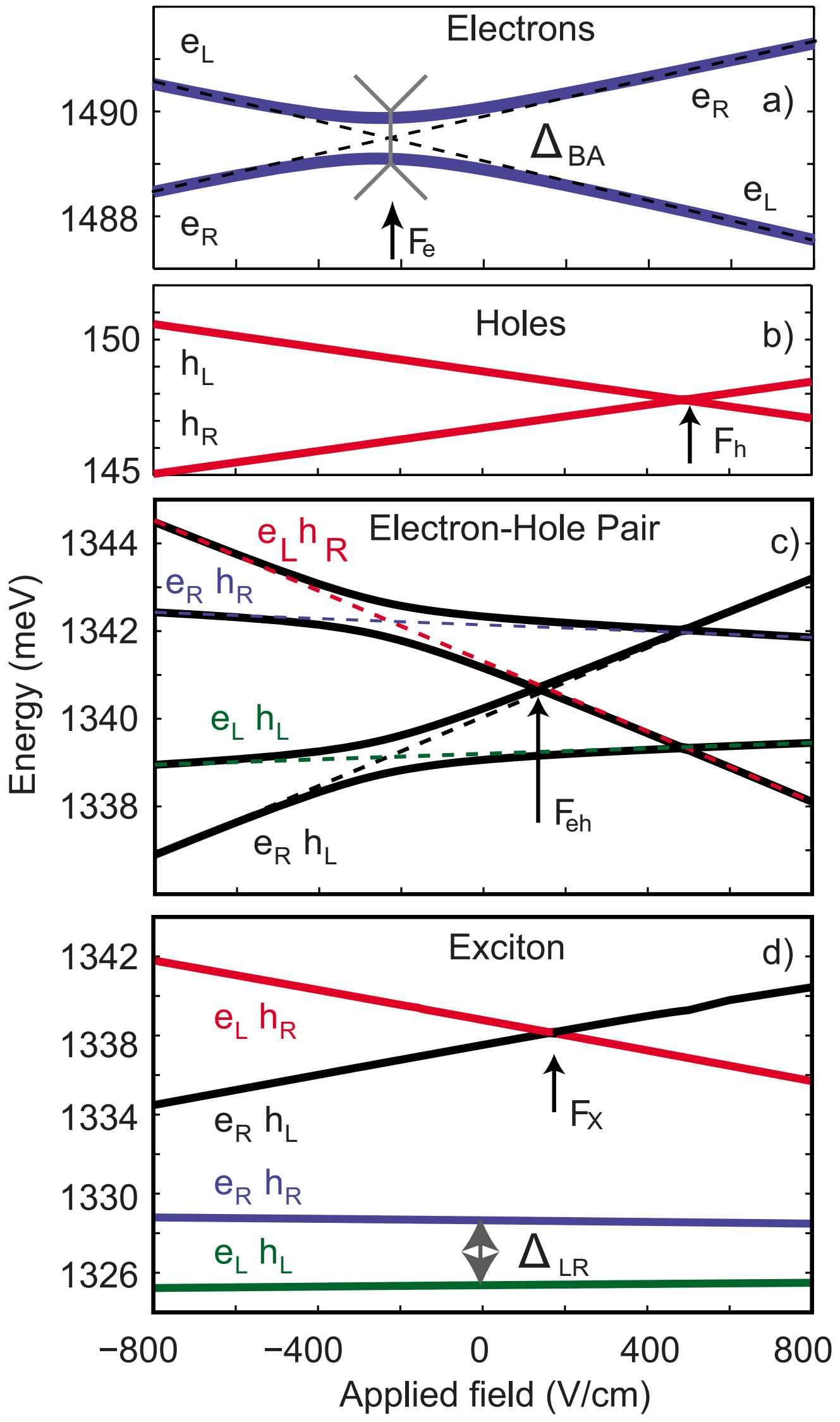

FIG. 2. (Color online) (a) Electron and (b) hole single-particle energies. Energies (c) of the uncorrelated electron-hole pair and (d) of the correlated exciton.

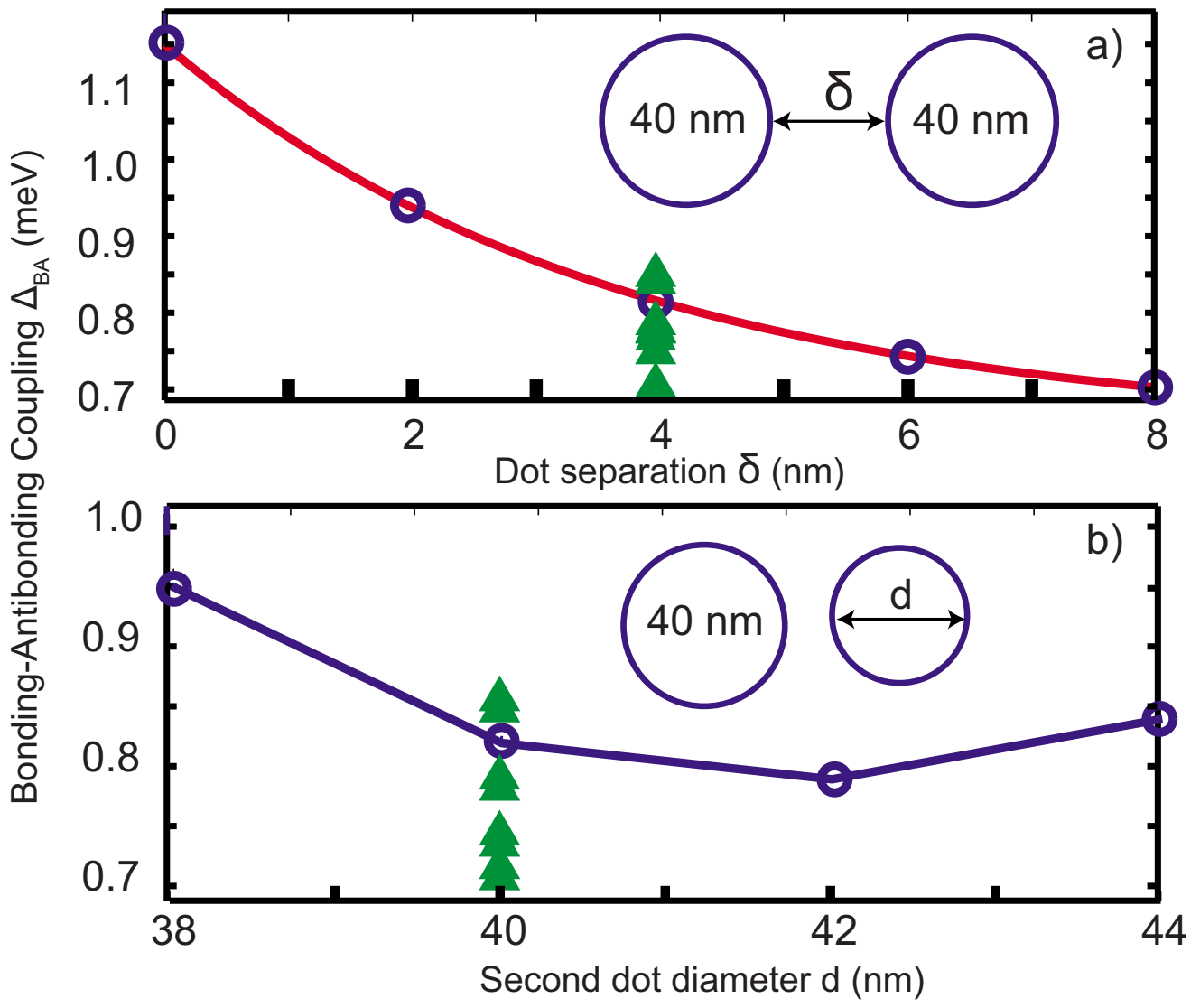

FIG. 3. (Color online) Bonding-antibonding splitting for the structure define in Figs. 2(a) and 2(b) for (a) varying interdot separation $\delta$ and (b) varying second dot diameter $d$. The triangles show energy separation of the two excitons for ten QDMs with the same structural parameters [Figs. 1(a) and 1(b)] but constructed from different random-alloy realizations.

### B. Dots with unequal diameters

We have calculated the full field dependence keeping the diameter of one dot at 40 nm and varying the diameter of the second dot as 38, 40, 42, and 44 nm, akin the experimental fluctuations. In Fig. 3(b), we see that the electronic coupling is mainly unaffected by the different dot sizes. This weak dependence can be explained by the large diameter of the dots compared to the exciton Bohr radius.

### C. Effect of the random-alloy realization

Keeping the structural parameters fixed, according to Figs. 1(a) and 1(b), we calculated ten QDMs made of different randomly generated alloys (but keeping the average composition fixed). The results are shown in Fig. 3 as filled green triangles. We obtain couplings between 0.65 and 0.85 meV in a range similar to the variation we obtain by changing the morphology of the dots.

These calculations lead to the conclusion that variations in diameters, as measured experimentally, do not significantly influence the results. The results remain qualitatively the same and even quantitatively, only small variations could be obtained. In general, two structures with identical size, shape, and composition but generated with different random alloys show variations in the same range as the variations we obtained by the changes in size and dot separation. The coupling of electronic states in this structure was not expected *a priori* because the dot centers are separated by 44 nm, an order of magnitude more than in the well-studied system of vertically coupled quantum dots.[16–18]

## VI. EXCITON FORMATION AND OPTICAL PROPERTIES

In Fig. 2(c), we plot the energy of the electron-hole pair, neglecting all few body effects (Coulomb interaction and correlations) as solid lines. The dashed lines are the results obtained after projecting the single-particle results onto a dot-localized basis, where the particles are either on the left or the right dot. We recognize the *indirect* pairs $e_Lh_R$, $e_Rh_L$, crossing at $F_{\rm eh}=120$ V/cm with a strong-field dependence. The *direct* pairs $e_Rh_R$, $e_Lh_L$, are only weakly dependent on field. When Coulomb attraction is included, the two direct pairs are shifted down by $\simeq 12$ meV while the indirect pairs are mainly unshifted. This consideration is sufficient to qualitatively understand the results for the correlated exciton given in Fig. 2(d). The lower two states belong to the direct excitons across the entire field range studied. The indirect excitons are at higher energy and cross at $F_{\rm X}=180$ V/cm. The theoretical absorption results are given in Fig. 4(a). They feature, not surprisingly, two bright states, almost independent of the field, corresponding to the direct excitons in Fig. 2(d) and are separated by $\Delta_{LR}$. The indirect excitons are entirely dark since there is vanishing overlap between the wave functions. This is in contrast to vertically coupled dots, where the indirect states acquire some oscillator strength through a significant electron-hole overlap. The almost vanishing slopes of the $e_Lh_L$ and $e_Rh_R$ exciton branches, are a consequence of the small in-plane permanent electric dipole of the $e_Lh_L$ and $e_Rh_R$ excitons. An excitonic dipole leads to the linear term in the quantum-confined Stark effect and to two opposite slopes for $e_Lh_L$ and $e_Rh_R$, and to their eventual crossing. In an isolated QD, the shift of the electron from the hole wave functions *in plane* can only be the result of different random alloys and is very small. For our QDM, the basin attracts the electron to the center of the structure while the holes remain located in the center of each dot [see Fig. 1(c)]. This leads to the small permanent dipole reflected in the slightly different slopes of $e_Lh_L$ and $e_Rh_R$. In the case of vertically coupled QDs, the composition gradient measured along the growth direction leads to a significant permanent electric dipole in growth direction and to rather strongly tunable direct excitons.

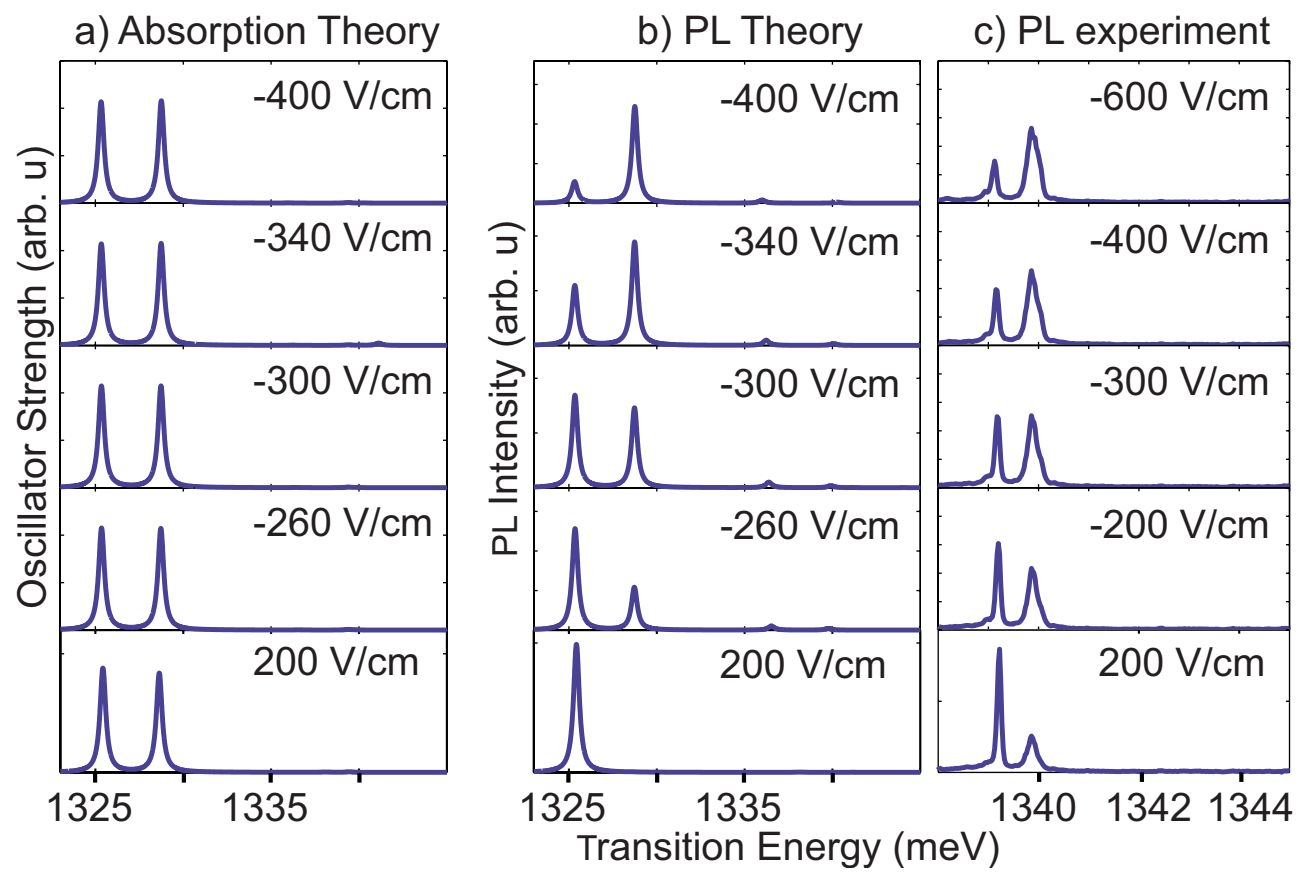

FIG. 4. (Color online) Theoretical results for (a) the absorption and (b) the PL. (c) Experimental PL results obtained under nonresonant excitation.

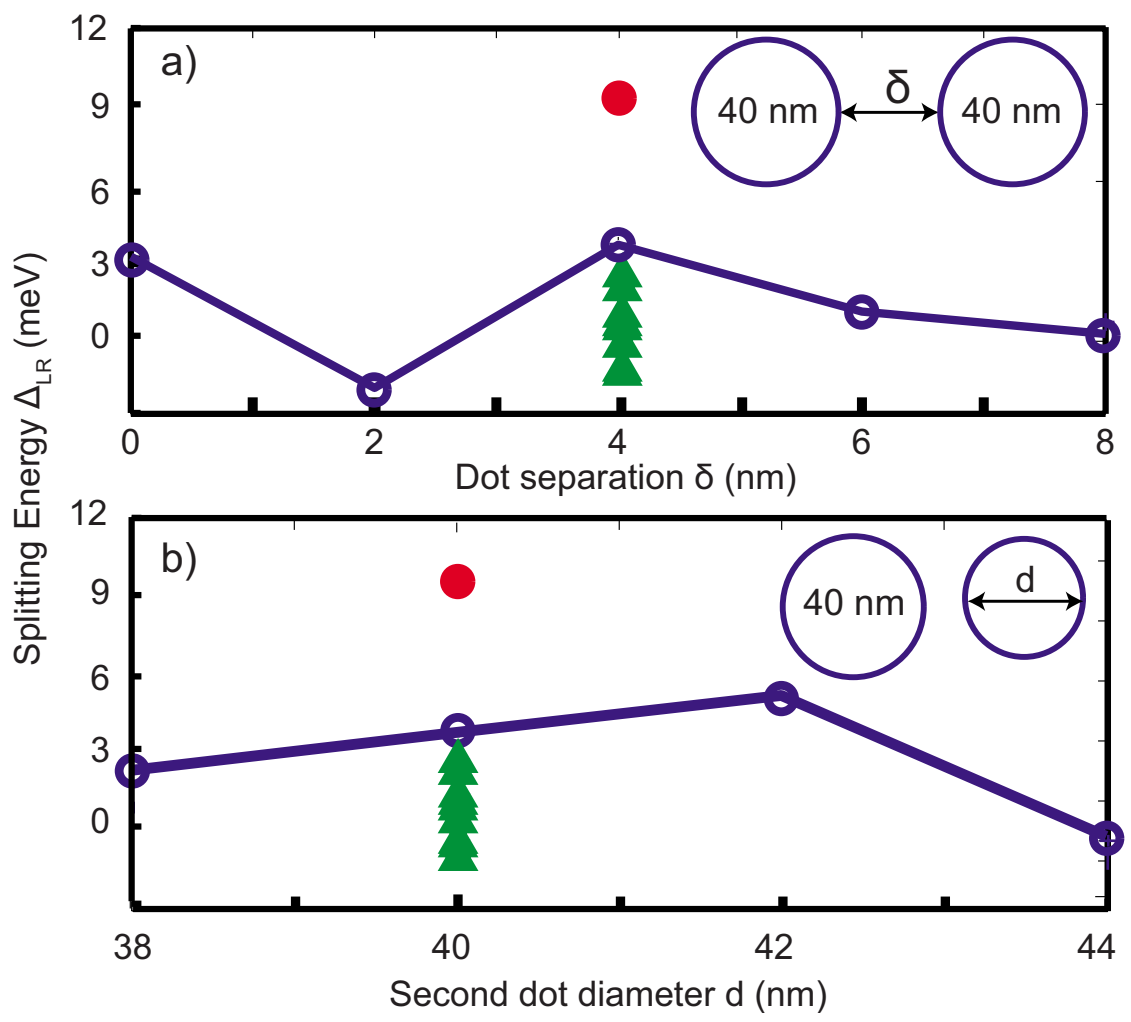

FIG. 5. (Color online) Energy separation of the two excitons $\Delta_{LR}$ for the structure define in Figs. 2(a) and 2(b) for (a) varying interdot separation $\delta$ and (b) varying second dot diameter $d$. The triangles show separations for ten QDMs with the same structural parameters [Figs. 1(a) and 1(b)] but constructed from different random realizations. The filled circles represent a structure similar to Figs. 1(a) and 1(b) but with two dots with a height difference of one monolayer.

### A. Separation between $e_Lh_L$ and $e_Rh_R$, $\Delta_{LR}$

The energetic separation between the two bright states at zero field is mainly given (neglecting small correlation effects) by the difference in the exciton energy of the left and the right dot, and does not indicate coupling. From this point of view, we would expect a large range of possible separations. In Fig. 5, we show that, if we vary the interdot separation or allow for some size mismatch between the dots, $\Delta_{LR}$ varies only in a narrow range between 0 and 5 meV. Our results for dots of identical overall shape and composition but made of differently generated random alloys are shown as green triangles in Fig. 5 and show a very similar range of $\Delta_{LR}$. However, if we change the height of one of the dot by only one monolayer, the separation jumps to a value of 9 meV, as shown in Fig. 5 as filled red circles. Experimentally, separations $\Delta_{LR}$ between 0.3 and 2.5 meV have been measured in excellent agreement with the calculations where both dots have the same height.

### B. Anticrossings

At larger electric fields (in our case at 2500 V/cm), $e_Lh_L$ and $e_Rh_R$ eventually cross (hole tunneling being negligible as well as Förster coupling on grounds of the large interdot separation) in contrast to the case of vertically coupled dots, where the absorption peaks should undergo anticrossings due to electron and hole tunneling.[16] Also, we do not expect anticrossings related to hole tunneling alone, which can be observed in vertically coupled dots.[17,18] However, anticrossings related to the electron tunneling should occur when the $e_Lh_L$ and the $e_Rh_L$ ($e_Rh_R$ and $e_Lh_R$) branches cross, at $-3100$ V/cm (2500 V/cm) in our case. These avoided crossings were the focus of previous theoretical studies.[19,20] With the present experimental setup, these fields cannot be achieved but may become available in the near future. In general, since the indirect excitons have very low oscillator strength, the spectral range in which the anticrossing may be observed will be narrower than in the case of vertically arranged QDMs.

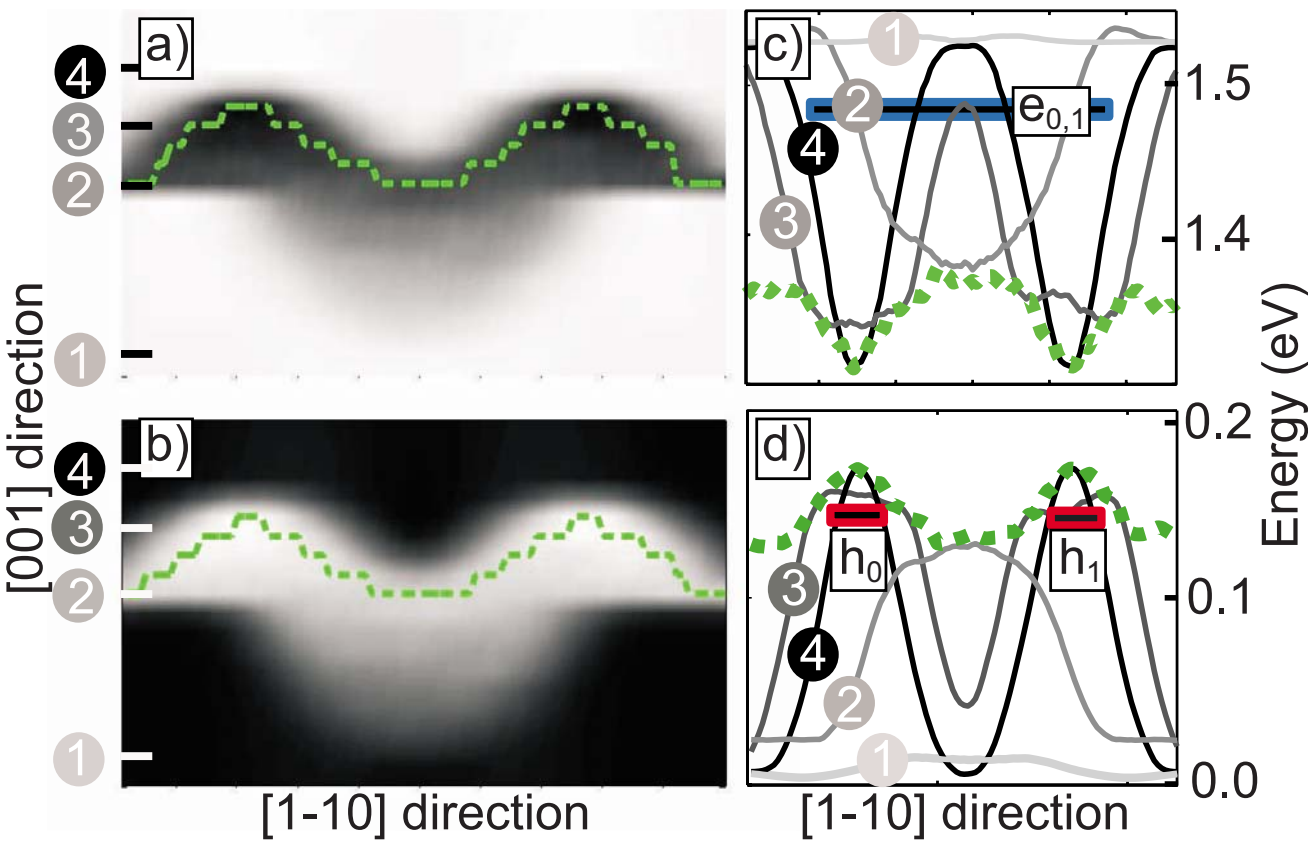

FIG. 6. (Color online) Confinement potential for the electrons (a) and (c) and the holes (b) and (d) in the (110) plane (a) and (b) and along the $[1\bar{1}0]$ direction (c) and (d) for different heights (positions along [001]) labeled as 1, 2, 3, 4. The green dashed lines in (a) and (c) follows the maximum (minimum) potential energy in the plane. The dashed green lines in (c) and (d) correspond to the energy values of the path given by the green lines in (a) and (b).

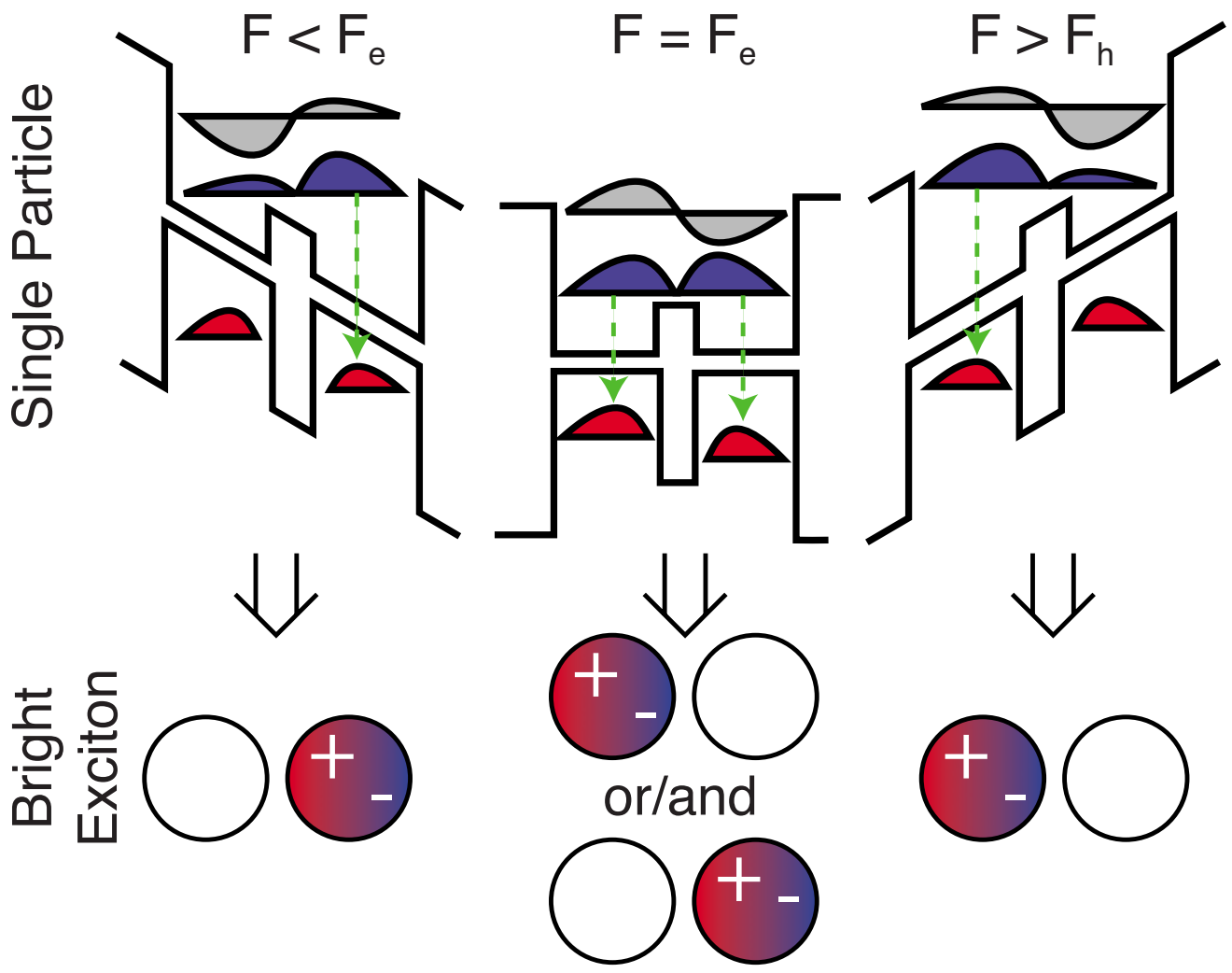

FIG. 7. (Color online) Schematic representation of the emission process. The hole states are well confined in either of the dots while the electron states form delocalized bonding and antibonding state. Following our dynamical model, where the upper electron state is effectively emptied, only the lower state is occupied. In the lower part of the figure, we show the exciton state ensuing from the single-particle states. The exciton is localized inside the left dot for $F<F_e$, inside the right dot for $F>F_h$, and can be localized on either of the dots in the intermediate range of applied fields. The words "and/or" mean that in the case of a single exciton recombination the photon will be emitted from one or the other dot. In the case of PL, there will be a statistical average of emission from the left *and* the right dot.

## VII. PHOTOLUMINESCENCE AND DYNAMICAL PROCESSES

The calculation of PL involves dynamical processes. While this is always true, in single quantum dot experiments, the absorption and emission are tightly related and besides a temperature effect (only the lowest exciton states are occupied in PL), the qualitative picture of both processes are similar. In the present case, the dynamical processes have profound consequences leading to a qualitative different picture of absorption, nonresonant, and resonant PL. To understand the underlying processes, we start by plotting in Fig. 6 the confinement potential felt by (a) the electron and (b) the hole as a false color image in the (110) plane. The dashed lines in Figs. 6(a) and 6(b) connect the right with the left dot following the floor of the potential valley. The corresponding potential energies are shown as dashed lines in Figs. 6(c) and 6(d). Figures 6(c) and 6(d) also show the confining potential energies along the $[1\bar{1}0]$ direction at different positions along [001] labeled as 1, 2, 3, 4. The single-particle energies of the electron (hole) states $e_{0,1}$ ($h_{0,1}$) are drawn as thick bars. The conclusion emerging from this picture is that the eigenvalues of $e_0, e_1$ in Fig. 6(c) are energetically above the "barrier" given by the potential-energy increase between both dots. In other words, if the electron would be seen as a classical particle it would not feel any barrier by moving from one dot to the other. The wave-function character and the delocalized nature of the quantum particle lead to a somewhat relaxed statement since the particle probes not only the valley floor but also its vicinity. We argue that the electron feels a very shallow potential barrier and sees the structure as one compound rather than two isolated dots. The situation for the holes is drastically different as they feel a steep potential barrier and are confined to each of the dots.

### A. Consequences of movable electrons and trapped holes

This important single-particle realization leads to the sketch given in Fig. 7, where the electron states are shown above barrier and the holes are well confined. The consequence for our dynamical model is as follows. If the electron and the hole enter the dot molecule as a bound exciton, they climb down the excitonic ladder, landing statistically either in the right ($e_R h_R$) or the left dot ($e_L h_L$). Both $e_R h_R$ and $e_L h_L$ exciton states are bright at all fields. If the hole comes first and occupies statistically either the left or the right dot and remains without the ability to tunnel, again, both $e_R h_R$ and $e_L h_L$ exciton states can be observed at all fields. In Fig. 4(c), we show the experimental results for a nonresonant excitation. These results are in direct contradiction with the scenarios described above and show a switching of the PL intensity from one to the other branch with the applied field. This behavior is, however, in excellent agreement with the theoretical results of Fig. 4(b) where we have assumed that the electron enters the structure first. In this case, the electron quickly relaxes to the lowest single-particle state $e_0$ (blue in Fig. 7) and the state $e_1$ (gray in Fig. 7) is effectively unoccupied. At finite electric fields this means the electron would be localized either on the left ($F>F_e$) or on the right ($F<F_e$) dot. When the hole enters the structure it forms the bright exciton $e_L h_L$ for $F>F_e$ and the bright exciton $e_R h_R$ for $F<F_e$, hence only one peak in PL is observed. At the electron tuning field $F=F_e$, the lowest electron state is distributed between both dots (bonding state) and both excitons $e_L h_L$ and $e_R h_R$ can be populated, hence two peaks are observed in PL. Consequently, the nonresonant PL experiment can be used to determine $F_e$, i.e., the single-particle electron "tuning" field from Fig. 2(a), of every investigated QDM.

Hence, this signal from switching of intensities gives us a direct insight into the localization delocalization of the single lone electron. It is interesting to note that an optical signal originating from an excitonic two-particle state, allows to draw conclusions about the single-particle nature of one of its constituents (the electron in our case). This is true although the exciton states around the switching of intensities [or equivalently the anticrossing of the single-particle electron states in Fig. 2(a)] are constituted of (mainly) two single-particle electron states. This is in contract to the case of anticrossings in excitonic energies (or PL lines),[17,21] where the anticrossing field does not reveal the tuning of the single-particle electrons or holes but is the field at which the direct and the indirect excitonic branches happen to (anti)cross.[16] This excitonic anticrossing point is the complex consequence of morphological conditions and is rather difficult to predict, also because of the many-body nature of the exciton. The switching of intensities still depends on the morphology but has a rather intuitive meaning: it happens at the field where the single-particle electron S states are aligned.

### B. Possible justification for an early electron capture

To assume an early electron capture is reasonable. Indeed, it is believed (see e.g., Ref. 22, and references therein) that a nonresonant optical excitation leads to the formation of hot electrons and holes. These must be allowed to cool for several picoseconds[23,24] prior to the formation of a hot exciton, and the eventual carrier capture by the QDs or the wetting layer, typically within a few picoseconds.[25–27] In the initial "single-particle" phase, the hot electrons and holes are rather independent. At a low excitation power, the processes of drift and diffusion toward the QDs, where electrons are faster than holes, are probably dominant. So an electron capture by the dots is expected prior to the hole capture. Moreover, if this diffusion motion occurs through the wetting-layer states, the heavier holes may be trapped by local potential fluctuations and slowed down compared to the lighter electrons. This effect was proposed to explain the dominance of trion states in recent experiments,[28] and may be an important factor in experimental setups involving lateral fields. The situation in vertically coupled QDMs is different with respect to the hole tunneling but the early electron capture is a likely process as well. Indeed, the occupation of the shallower (usually top) QD by electrons [in the notation of Scheibner *et al.*[21] $\left(\begin{smallmatrix} e_Be_T \\ h_Bh_T \end{smallmatrix}\right) = \left(\begin{smallmatrix} 01 \\ 00 \end{smallmatrix}\right)$] seems unobserved.[21] One possible explanation would be a fast electron capture followed by a tunneling into the lowest single-particle electron states [into the bottom dot, $\left(\begin{smallmatrix} 10 \\ 00 \end{smallmatrix}\right)$], a scenario similar to the one described here for the lateral QDMs.

## VIII. RESONANT EXCITATION

The model derived for the PL can be tested for the situation of resonant excitation, where the electron-hole pair is directly injected into the QDMs, energetically below the barrier band gap. This is a very powerful method since it eliminates all the intricacies of carrier drift/diffusion and exciton formation. Our understanding in this case can be represented by the double ladder given in Fig. 8(a). The rungs represent the calculated exciton energies and are separated in a left (right) ladder for exciton states where the hole is mostly localized in the left (right) dot. The resonantly created exciton is a direct exciton (only these are bright) of the type $e_Lh_L$ or $e_Rh_R$ at the energies ER1 and ER2. Since the hole is unable to tunnel between the dots, the exciton relaxation/cooling uses either one or the other ladder but not both. Consequently, resonant excitation leads to a single peak, independent of the applied field. In Fig. 8(b), we show the results for a nonresonant excitation similar to Fig. 4(c) but for another QDM, where the field is set close to the tuning point $F=F_e$ where both peaks are bright. In Figs. 8(c) and 8(d), we show the experimental results of two resonant excitations at the energies ER1=1361.9 meV and ER2 =1359.2 meV. Figure 8(c) shows a more intense low-energy peak while Fig. 8(d) shows a more intense high-energy peak. This supports our model, where ER1 excites a bright state in the left ladder and ER2 a bright state in the right ladder [the energetic position of the resonant excitations are given by ER1′ and ER2′ for the theoretical results of Fig. 8(a)] leading in each case to a single PL line. The remaining luminescence of the other peaks is attributed to additional nonresonant pump channels and the finite spectral width of the pulsed laser which allows for simultaneous excitation of different states.

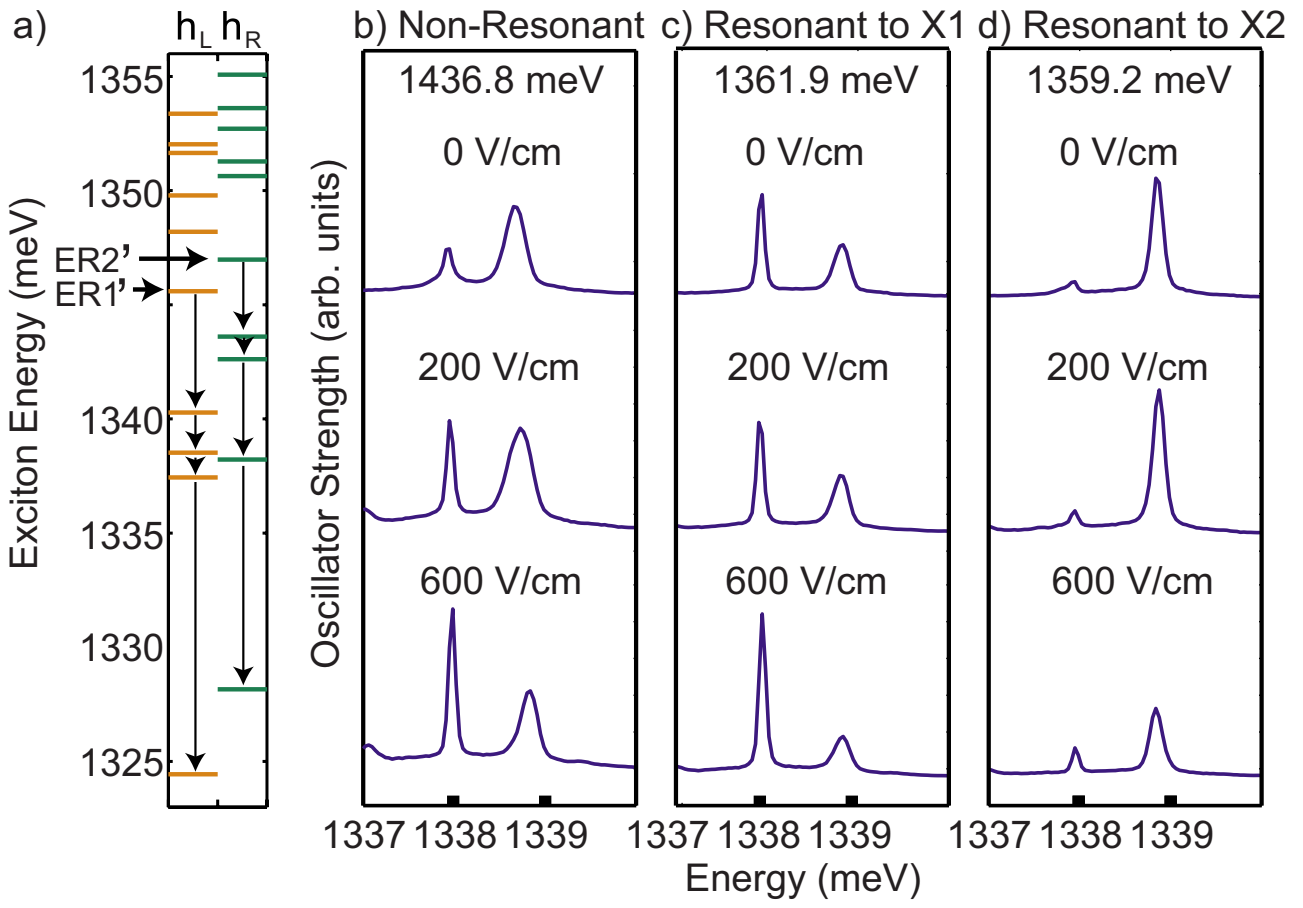

FIG. 8. (Color online) (a) Calculation of the correlated exciton states in the dot molecule. The exciton ladder has been split according to the dominant position of the hole. (b) Experimental PL results for nonresonant excitation at energy $E=1436.8$ meV. (c) and (d) Experimental resonant excitations at energy ER1=1361.9 meV and ER2=1359.2 meV.

## IX. DISCUSSION AND CONCLUSIONS

The aforedescribed situation where the electron position can be controlled using a lateral electric field could be exploited to use single and individually prepared electrons within the QDM as a charge or spin qubit. With the ability to have such electrically prepared and positioned single resident electrons in a QDM, a circularly polarized light field that is in resonance with the respective charged exciton trion ground- or excited-state transition could be used to prepare or manipulate the electron-spin state. Sequences of microwave and/or optical pulses of well-defined polarization, width, and amplitude ($\pi/2$, $\pi$, etc.) could be used for precise state initialization, manipulation, (e.g., spin flip or rotation), and readout. Two qubit gates may be realized using electrons in a lateral QDM, e.g., the charge/spin state in one dot (target) could be manipulated depending on the state of the other dot (control), resulting in a CNOT or CROT gate.

Concluding, we showed that by the combination of an atomistic many-body approach, knowledge of the detailed morphology of the QDMs, nonresonant, and resonant PL experiments, we reach a thorough understanding of the underlying processes involved in the optical experiments. We highlight the importance of electronic coupling in lateral dot molecules fostered by the presence of an In-rich basin connecting the dots from below. This leads to a peculiar confinement situation with regions of type I band alignment close to a coupling region on the verge to type II. Beyond the static picture, we find strong evidence, backed up by PL measurements, for a dynamical model ensuing from the lack of potential barrier felt by the electron in opposition to the decoupled holes. This model leads to a qualitatively different behavior for absorption, emission after nonresonant excitation, and emission after resonant excitation. In absorption, two peaks are expected at all fields. Only one peak is expected and observed under resonant excitation, due to the existence of two distinct exciton relaxation paths. The PL under nonresonant excitation is expected to show switching from one exciton branch to the other with the applied field, which coincides with experiment. This very rich dynamics is a consequence of the unusual confinement potential felt by the electron and holes and by the difference in the electron and hole capture mechanism. One of the striking features of the present study is the relative simplicity of the exciton manifold [Fig. 2(d)]. The high level of confidence about its accuracy is based on a demonstrated relative insensitivity about details of the morphology and the direct comparison with PL experiments under resonant and under nonresonant excitations. This allows us to draw conclusions about dynamical processes, such as the early electron capture, usually too complex to disentangle from an often uncertain and complex excitonic structure. Having the ability to manipulate the position of the significantly coupled electrons within the QDM leaves it not only as a wavelength tunable single-photon source but also as a potential building block for quantum gates based on single charges and spins. In order to achieve this, regimes of coherent coupling within the QDM have to be investigated, such as the resonant tunneling of an electron.

## ACKNOWLEDGMENTS

Work at the University of Stuttgart was financially supported by the DFG via SFB/TR21 and FOR730.